\documentclass[12pt]{iopart}

\usepackage{graphicx}
\usepackage{epsfig}
\usepackage{subcaption}
\usepackage{lineno} 
\expandafter\let\csname equation*\endcsname\relax
\expandafter\let\csname endequation*\endcsname\relax
\usepackage{amsmath}
\usepackage{hyperref}

\begin{document}

\title[Filtering hits for speeding up online track reconstruction at hadron colliders]{Filtering hits for speeding up online track reconstruction at hadron colliders}

\author{Andrea Coccaro$^{1}$, Carlo Schiavi$^{1,2}$, Alessandro Zaio$^{1,2}$}
\address{$^{1}$INFN, Sezione di Genova, 16146 Genova, Italy}
\address{$^{2}$Department of Physics, Università di Genova, 16146 Genova, Italy}

\ead{andrea.coccaro@ge.infn.it, carlo.schiavi@ge.infn.it, alessandro.zaio@ge.infn.it}

\vspace{10pt}

\begin{abstract}
Collider experiments are equipped with trigger systems that rapidly inspect the physics content emerging from collisions to decide whether the resulting products are worth saving for later analysis. One crucial aspect for analyzing the final states originating from the collisions is to process the information produced by charged particles in the innermost detectors to reconstruct the corresponding trajectories. This task is a challenge for the experiments running at the Large Hadron Collider (LHC) at CERN because of the large number of secondary collisions per bunch crossing, the so-called pile-up vertices, giving rise to extremely high hit occupancies in the detector layers close to the beam line.
Reconstructing tracks is a combinatorial problem and its processing time strongly depends on the average pile-up per event. The future accelerator-complex upgrade to the High-Luminosity LHC, implying even higher detector occupancies, will result in a considerable growth of the computational cost of the current trigger strategies. To face this issue, a new technique for assisting track reconstruction by filtering out unnecessary detector information is presented and characterized in this work. The algorithm is based on a convolutional-neural-network architecture which can be easily deployed on accelerator cards. The impact of this approach is assessed and future prospects are also discussed.
\end{abstract}

\vspace{2pc}

\section{Introduction}
Experimental particle physics studies the fundamental particles and forces that make up the universe by using high-energy accelerators, such as the LHC~\cite{ref:LHC}, and detectors, such as ATLAS~\cite{ref:ATLAS-Paper} and CMS~\cite{ref:CMS-Paper}. At the LHC, interesting events from the proton-proton collisions are rare and the ATLAS and CMS experiments are equipped with sophisticated trigger systems for collecting and filtering detector data from the initial collision rate of $\mathcal{O}$(40 MHz) to $\mathcal{O}$(1 KHZ) for further physics inspection and analysis~\cite{ref:ATLAS-Trigger-Run2,ref:ATLAS-Trigger-Run3,ref:CMS-Trigger}.

Events are selected by a two-level trigger system. While the first level operates using custom hardware connected to the calorimeters and the muon spectrometer, the second software-based trigger also processes information from the highly-granular inner tracking detector, enabling a more comprehensive suite of particle reconstruction and identification tools. The inner tracking detector records hits from charged particles and its data volume is extremely large. The reconstruction of tracks from these hits represents one of the most CPU-time consuming tasks operating at the trigger level~\cite{ATLAS:2021lws}. It is so computationally expensive because of the intrinsic combinatorial nature of the algorithm, and because of the high-density environments with many particles and detector layers.

When the LHC runs at its highest operation intensities, the mean number of simultaneous collisions per bunch crossing goes to~$\sim70$ but with the upcoming High-Luminosity LHC (HL-LHC) upgrades the instantaneous luminosity is expected to increase from 1 to $7.5\times10^{34}$\,cm$^{-2}$\,s$^{-1}$ and pile-up from the current 50--70 up to 150--200~\cite{ref:HL-LHC}. At the HL-LHC, the current track reconstruction approach at the trigger level becomes prohibitive because of the higher detector occupancy and the higher number of readout channels of the upgraded detectors~\cite{ref:ATLAS-PhaseII,ATLAS:2024rnw}. New proposals to mitigate the computational timing cost, also based on machine-learning advancements, recently appeared in literature~\cite{Duarte:2020ngm,DeZoort:2021rbj,Bocci:2020pmi}. In addition, heterogeneous computing offers an interesting path of further investigation, in particular both GPU-based and FPGA-based accelerations are a promising option for the data acquisition systems of future particle-physics experiments~\cite{ATLAS-HL-LHC-Computing,CMS-HL-LHC-Computing,Coccaro:2023nol,Soybelman:2024mbv}.
In this context, the work presented in this manuscript introduces a novel approach for filtering the hit information prior to any reconstruction task. The filter acts on the reconstructed hits by classifying them according to the probability of being associated to a particle originating from the hard-scatter primary vertex or from a pile-up collision or other background sources. The usefulness of this approach is demonstrated by characterizing its performance on synthetic data mimicking the LHC environment and the ATLAS tracking detector. It is demonstrated that the number of hits entering a track reconstruction algorithm can be substantially reduced, and so the processing-time footprint, without sacrificing the physics output. Moreover, the adopted algorithm, based on a rather simple convolutional neural-network architecture, allows an easy deployment on both GPU and FPGA accelerator cards, enabling fast inference on heterogeneous computing to be further explored. The data generation is described in Section~\ref{Sec:Data}, the filtering algorithm and its performance are presented in Section~\ref{Sec:Algorithm}, while additional robustness studies are detailed in Section~\ref{Sec:Robustness}.

\section{Synthetic dataset}\label{Sec:Data}
Synthetic data is generated with a custom event generator simulating a cylindrical detector consisting of eight concentric layers. Its geometry is based on the barrel part of the inner detector of the ATLAS experiment, and with the four layers matching the pixel detector and the outer four layers the strip detector. A solenoidal magnetic field, aligned with the beam axis and as in the ATLAS experiment, is also emulated. A right-handed coordinate system is adopted with the origin at the nominal interaction point (IP) in the center of the detector and the $z$-axis along the beam pipe. The $x$-axis points from the IP to the center of the LHC ring while the $y$-axis points upwards. The azimuthal angle $\varphi$ is defined around the $z$-axis and the polar coordinates $(r,\varphi)$ in the transverse plane.

Experimental resolution effects are taken into account by smearing the $\varphi$ and $z$ coordinates of the simulated hits taking into account the precision of the emulated detector sensors, while the $r$ coordinate is assumed to be known perfectly. The applied gaussian smearing values are summarized in Table~\ref{tab:smearing}.

\begin{table}[t!]
\centering
\begin{tabular}{||l|c|c||l|c|c||}
\hline
pixel detector & r$\sigma(\varphi)$ [$\mu$m] & $\sigma(z)$ [mm] & strip detector & r$\sigma(\varphi)$ [$\mu$m] & $\sigma(z)$ [mm] \\
\hline
layer 0 & 50 & 0.25 & layer 4 & 130 & 2 \\
layer 1 & 50 & 0.4 & layer 5 & 130 & 2 \\
layer 2 & 50 & 0.4 & layer 6 & 130 & 2 \\
layer 3 & 50 & 0.4 & layer 7 & 130 & 2 \\
\hline
\end{tabular}
\caption{Standard deviations of the gaussian smearing applied to the generated hits in the simulated datasets. These values match the spatial dimensions and the typical related resolutions of the active sensors of the pixel and strip detectors of the ATLAS experiment \cite{ATLAS:IBL_TDR, ATLAS:Pixel_TDR, ATLAS:SCT_precision}.}
\label{tab:smearing}
\end{table}
\begin{table}[t!]
\centering
\begin{tabular}{|l|c|c|c|c|c|c|c|c|}
\hline
Layer index & 0 & 1 & 2 & 3 & 4 & 5 & 6 & 7 \\
\hline
Radius [mm] & 34.1 & 54.2 & 95.0 & 131.0 & 312.0 & 381.0 & 452.0 & 520.0 \\
\hline
Avg.\ number of hits & 100 & 104 & 98 & 92 & 60 & 54 & 49 & 46 \\
\hline
Avg.\ number of noise hits & 6 & 11 & 19 & 26 & 6 & 8 & 9.0 & 10 \\
\hline
Hit density [$10^{-4}$\,mm$^{-2}$] & 3.9 & 2.5 & 1.4 & 0.9 & 0.2 & 0.2 & 0.1 & 0.1 \\
\hline
Noise density [$10^{-5}$\,mm$^{-2}$] & 2.2 & 2.7 & 2.7 & 2.7 & 0.27 & 0.27 & 0.27 & 0.27 \\
\hline
\end{tabular}
\caption{The average hit counts and superficial densities for each layer of the simulated detector are shown, for both pileup and noise hits.The superficial densities are computed over the simulated acceptance window ($\Delta\phi = 0.2$\,rad, $\Delta z = 1200$\,mm).}
\label{tab:hit_density}
\end{table}

\begin{figure}[htbp]
\centering
\includegraphics[scale=0.7]{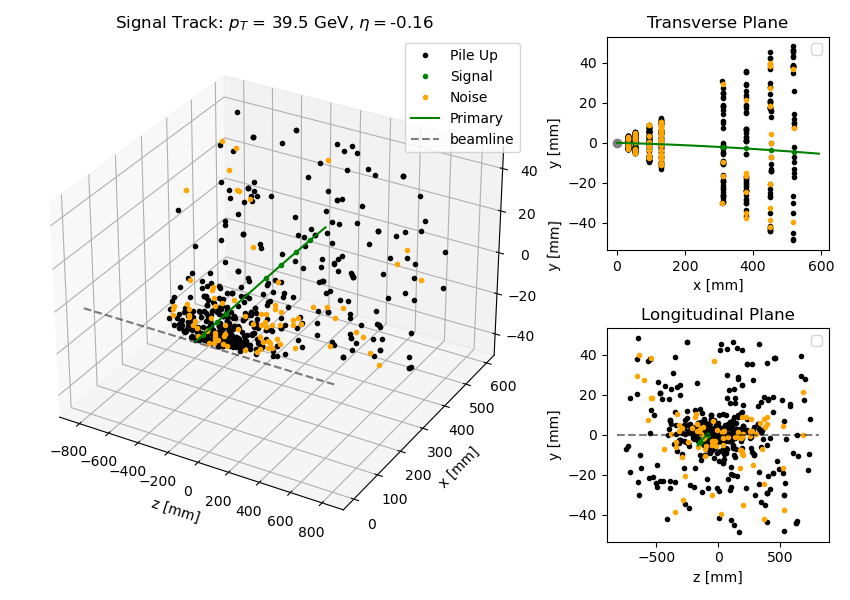}
\caption{Three-dimensional display of a generated event and projections in the $(x,y)$ and $(y,z)$ planes. The event contains a single high-momentum track with $p_{\text{T}}=39.5$\,GeV and a detector hit density due to secondary collisions from an event with a pile-up of 25. The hits are colored according to their truth nature: black for hits from pile-up particles, orange for hits from random noise and green for hits from the high-momentum signal particle. The beamline, in the three-dimensional and $(y-z)$ plane, or the beamspot, in the $(x,y)$ plane, is also shown in gray. For a better visualization, the green hits are also connected with a green line. More details on the data generation are in the text.}
\label{fig:event_display}
\end{figure}

Signal particles are generated as single charged particle tracks crossing all eight detector layers and with a transverse momentum $p_{\text{T}}$ between 20 and 50\,GeV. The 20\,GeV lower bound in the signal tracks corresponds to the typical selection of an isolated muon at the trigger level and hence is considered a representative benchmark for standard single-lepton trigger strategies.
To emulate a realistic detector environment, background collisions are generated with realistic distributions for the $p_{\text{T}}$ and the number of pileup tracks, and are then overlayed onto the signal track.
The background generation is tuned to emulate the per-layer hit density observed in simulated ATLAS Monte Carlo samples of semileptonic $t\bar{t}$ events for different pile-up settings~\cite{ATLAS:2024rnw}.
Particles emerging from pile-up vertices are generated in the full solid angle, but then only hits in regions of interest corresponding to $|\Delta\varphi|<0.1$ with respect to the generated signal particle are processed. The momentum distributions of pile-up particles follow the expected distribution according to public results~\cite{ATLAS:2016zkp}.
The primary and pile-up vertices are simulated according to the typical beam spot size, also along the $z$-axis direction.

Furthermore random noise hits are also generated across all layers to further challenge the hit-filtering task. The noise hits are generated uniformly across the $\varphi$ and $z$ coordinates, and the superficial hit-densities for each layer are specified in Table~\ref{tab:hit_density}. The layer index, used in the Table, ranges from 0 to 7 and identifies the detector layers from the innermost to the outermost and is in a one-to-one correspondence with the radius of the cylindrical coordinates.

An example of a generated event is depicted in Figure~\ref{fig:event_display}. In the figure, a high-momentum particle in the primary collision is generated together with 25 secondary collisions. The simulated number of pile-up collisions, for the events in the corresponding sample, is extracted using a Gaussian distribution with average 25 and width 5. Dots represent the hits and both a tree-dimensional view and two different two-dimensional projections are displayed. The dots come in different colors according to the scheme outlined in the caption of Figure~\ref{fig:event_display}. 

\begin{figure}[t!]
\centering
\includegraphics[scale=0.7]{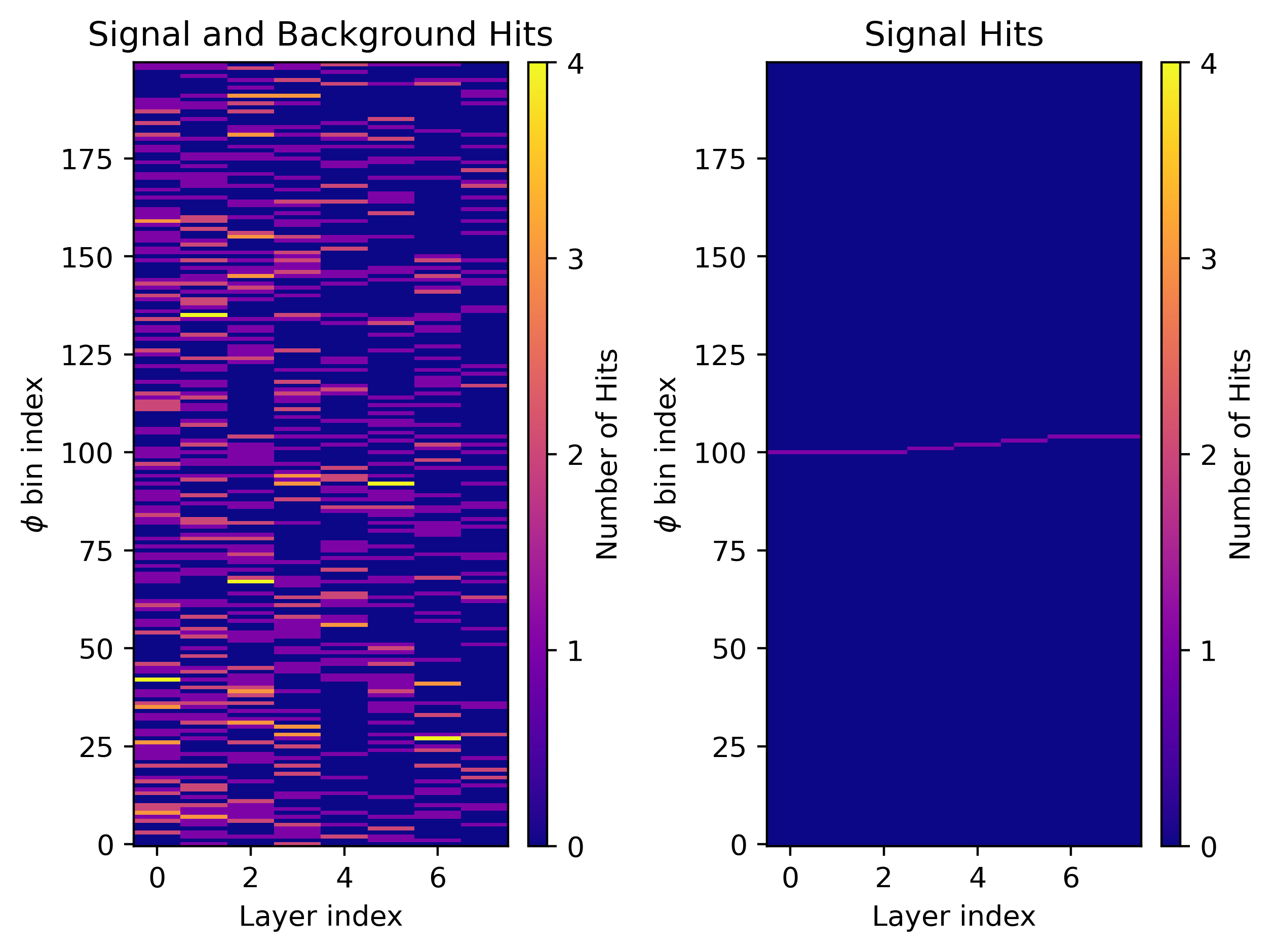}
\caption{Display of the hitmap obtained from the two-dimensional projection of a toy-event in the $(\varphi \text{ bin index}, \text{layer index})$ coordinates. On the left we have the hitmap for both signal and background while on the right only the signal is shown.}
\label{fig:event_image}
\end{figure}

The simulated events are then processed with a change of representation from the three-dimensional $(x,y,z)$ coordinates to the two-dimensional image-like $(\varphi \text{ bin index},\text{layer index})$ ones. Such manipulation creates a dataset of images, suitable for training a machine-learning algorithm with a convolutional neural-network architecture, as detailed in Section~\ref{Sec:Algorithm}.
The creation of two-dimensional images is effectively a compression, hence some discriminating information for filtering hits is inevitably lost. A possible outlook of the study presented in this manuscript is an algorithm manipulating three-dimensional images. This is certainly a desirable improvement and is left for future studies.
The image-like representation of the event in Figure~\ref{fig:event_display} is depicted in Figure~\ref{fig:event_image}. The binning on the $\varphi$ axis shown in Figure~\ref{fig:event_image}, is set to $0.001~\text{rad}$, corresponding to $r\varphi = 34$\,$\mu$m in the first layer and $\sim 500$\,$\mu$m in the last layer. The bin content corresponds to the number of hits, from signal or background sources, registered in that particular layer and width in the $\varphi$ direction. In this way, the information along the collapsed $z$-axis is, at least, partially encoded. 
Different bin widths in the $\varphi$ direction are explored since a finer binning may better separate nearby tracks and hence influence the performance of the filtering algorithm. 
% The baseline adopted binning reflects the trade-off among, on one side, the algorithm performance and, the other side, practical constraints related to the dataset size and the number of trainable parameters.

\section{The filtering algorithm}\label{Sec:Algorithm}
\begin{figure}[t!]
\centering
\includegraphics[scale=0.11]{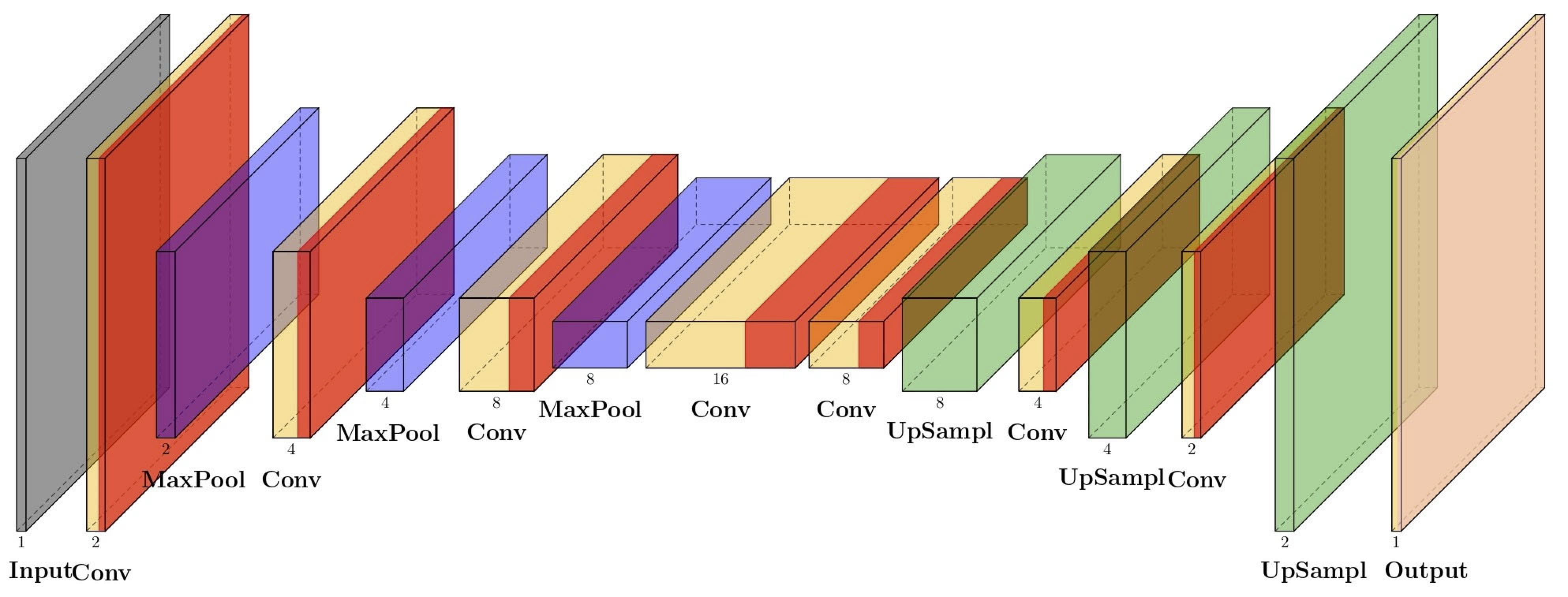}
\caption{Schematic representation of the architecture being developed for the filtering algorithm and based on a convolutional neural network.}
\label{fig:CNN}
\end{figure}

The filtering algorithm is designed to receive the data in the image-like form described in the previous section as input. The network architecture is inspired by the denoising autoencoders in Ref.~\cite{ref:reduce_dim, ref:denoise_ae, ref:segmentation}, which are designed to reconstruct a clean version of noisy input data by learning to classify between signal and noise at single-hit level. The architecture is an autoencoder, with a first encoder part to compress the information and a second decoder part to return to the original image dimensions, based on a convolutional neural network. The details of the actual architecture is depicted in Figure~\ref{fig:CNN}, where convolutional layers are interspersed with max-pooling layers in the encoder part while up-sampling layers constitute the decoder part. The Rectified Linear Unit (ReLU) activation function is used throughout the inner layers, while a sigmoid function is applied in the final output layer to allow the output to be interpreted as a probability. The adopted loss function is a weighted Binary Cross Entropy (BCE). The architecture of the model is consciously chosen as rather simple and light, to allow for an easy portability to different devices, including accelerator cards. This model comes with only $\sim$120k trainable parameters, which makes it particularly well-suited for fast-inference studies and low-memory footprint applications.

The dataset is composed of 1 million events with a single signal track of $p_{\text{T}}$ randomly extracted in between 20 and 50\,GeV plus the tracks from pile-up interactions and the additional detector noise. The training adopts the Adam optimizer and occurs for at most 500 epochs, and with early stopping implemented with a patience of 20 epochs. The trained network is then tested on an independently generated sample. The dataset is divided into 70\% for training, with the remaining 30\% split equally between validation and testing. 

\begin{figure}[t!]
\centering
\begin{minipage}[b]{0.49\textwidth}
\includegraphics[width=\textwidth]{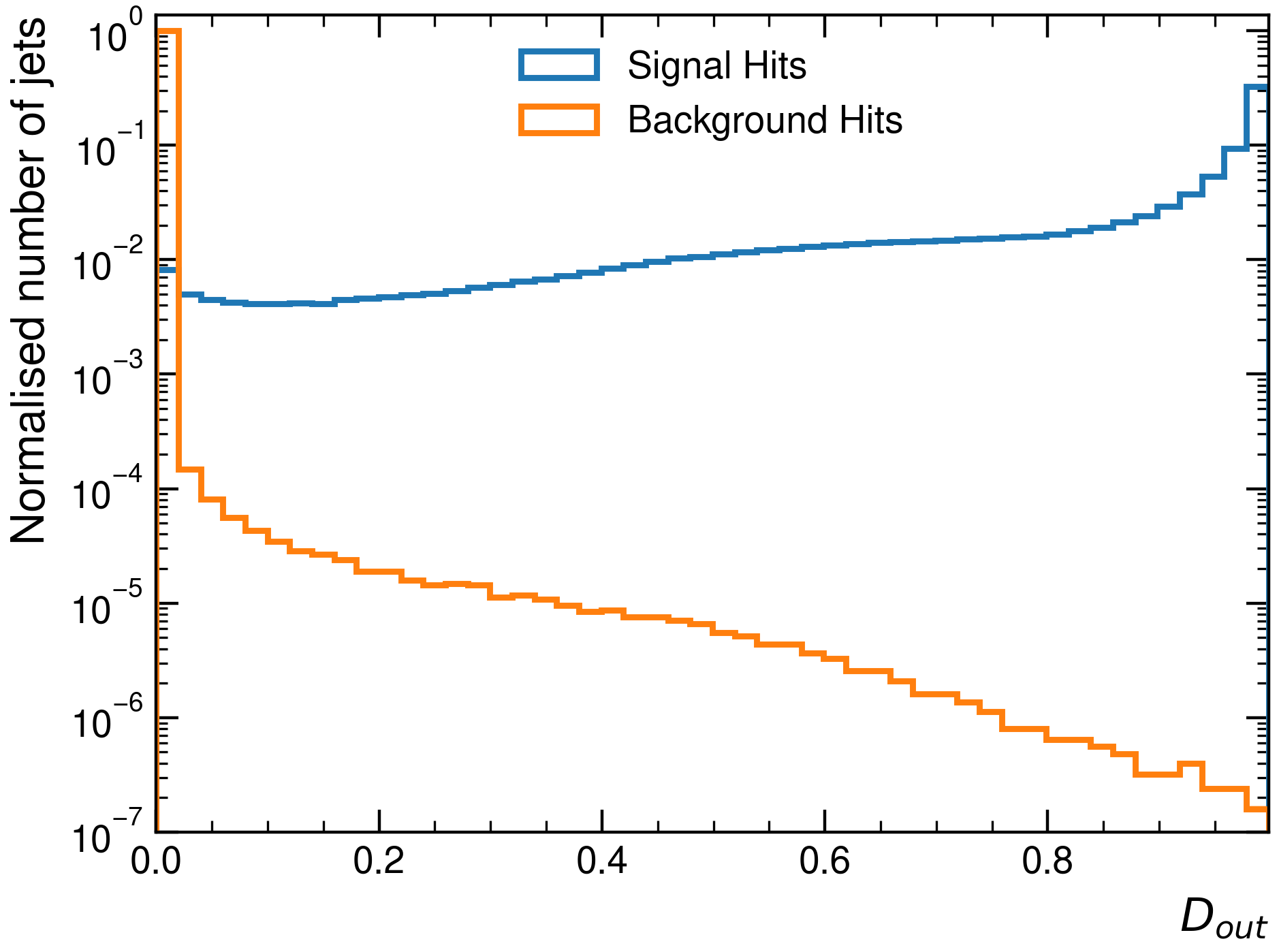}
\end{minipage}
\begin{minipage}[b]{0.49\textwidth}
\includegraphics[width=\textwidth]{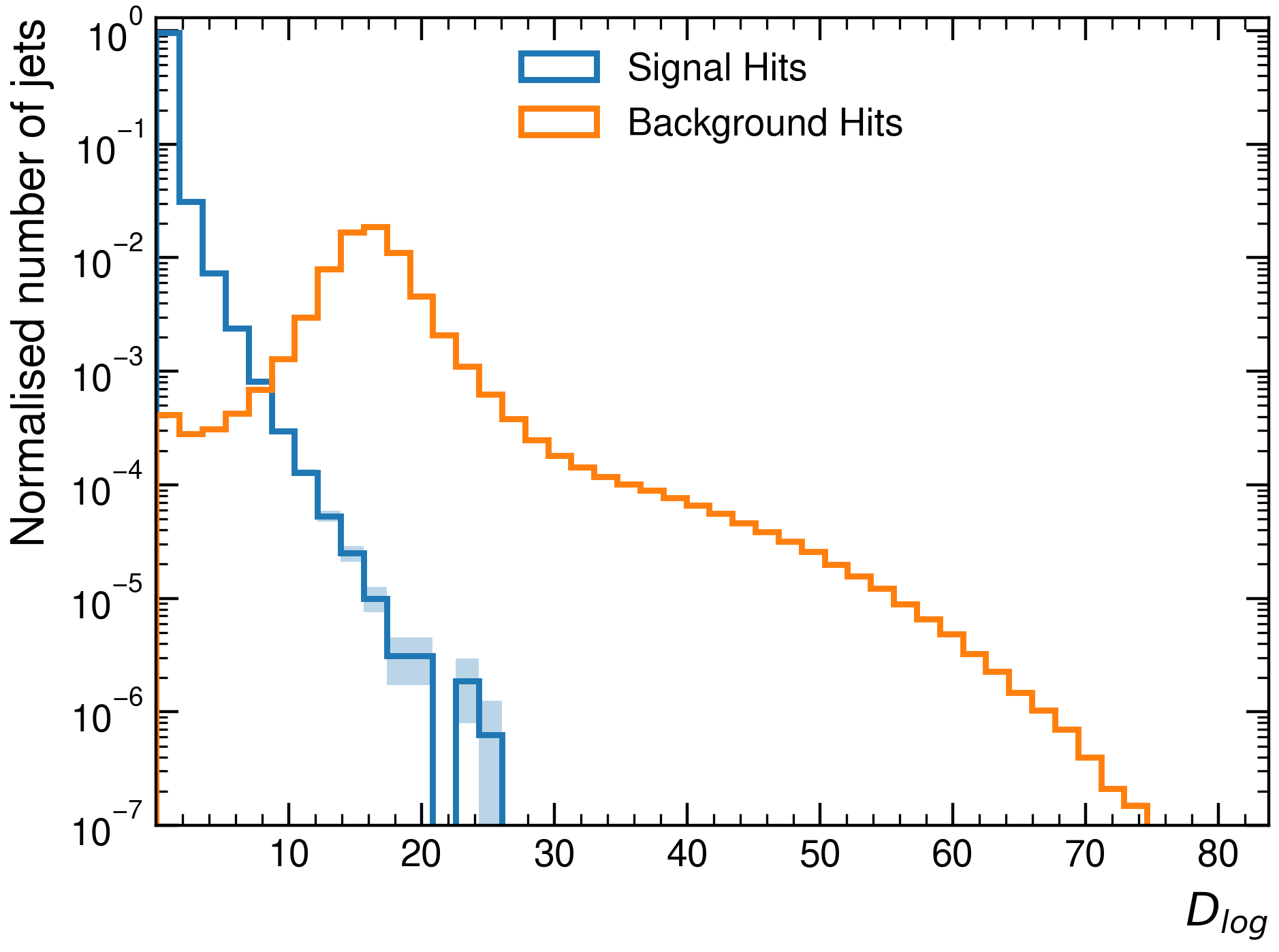}
\end{minipage}
\caption{The distribution of the raw $D_{\text{out}}$ output of the filtering algorithm using the model $K_{\text{start}}=128$, $K_{\text{end}}=4$ and the input dataset with a binning along the $\varphi$ direction corresponding to $\Delta\varphi=0.0002~\text{rad}$ is displayed on the left. The corresponding distribution of $D_{\text{log}}$, the scaling of the $D_{\text{out}}$, is displayed on the right.}
\label{fig:dicr_output}
\end{figure}

Before properly assessing the performance of the algorithm from a physics perspective a scan of the hyper-parameters is performed. In particular the $\varphi$ coordinate of the input data captures the discriminant information of the collapsed three-dimensional collisions hence the kernel dimension, in terms of pixels, along the $\varphi$ coordinate plays a crucial role. By defining $K_{\text{start}}$ and $K_{\text{end}}$ as the kernel dimensions respectively at the input layer and at the innermost layer different architectures are defined and a scan of all the various combinations is performed considering $K_{\text{start}} = \{32, 64, 128, 256\}$ and $K_{\text{end}} = \{1, 2, 4, 8, 16\}$. For all combinations the kernel dimension on the layer coordinate is kept to six and the depth of the autoencoder is varied such that at the next layer the dimension of the kernel on the $\varphi$ direction is halved until the dimension as in $K_{end}$ is reached. Additionally, to further characterize the importance of the $\varphi$ coordinate, three different datasets are created, each with a 1M event statistics, corresponding to a separation $\Delta\varphi$ in two contiguous bins of 0.001, 0.0005 and $0.0002~\text{rad}$. The various models with varying $K_{\text{start}}$ and $K_{\text{end}}$ are then trained on the three datasets with the same initial learning rate and early-stopping settings. The displayed result in Figure~\ref{fig:dicr_output} corresponds to the trained model with $K_{\text{start}}=128$, $K_{\text{end}}=4$ and $\Delta\varphi=0.0002~\text{rad}$. The shown distribution is the raw output, $D_{\text{out}}$, which is then scaled, for the performance studies shown in the later section, to $D_{\text{log}} = -\ln(D_{\text{out}})$, to avoid the large spike towards zero and to better characterize the performance for a wider range of selection efficiencies. The plot clearly shows that the neural network learned how to filter hits.

\begin{figure}[htbp]
\centering
\begin{minipage}[b]{0.45\textwidth}
\centering
\includegraphics[width=\textwidth]{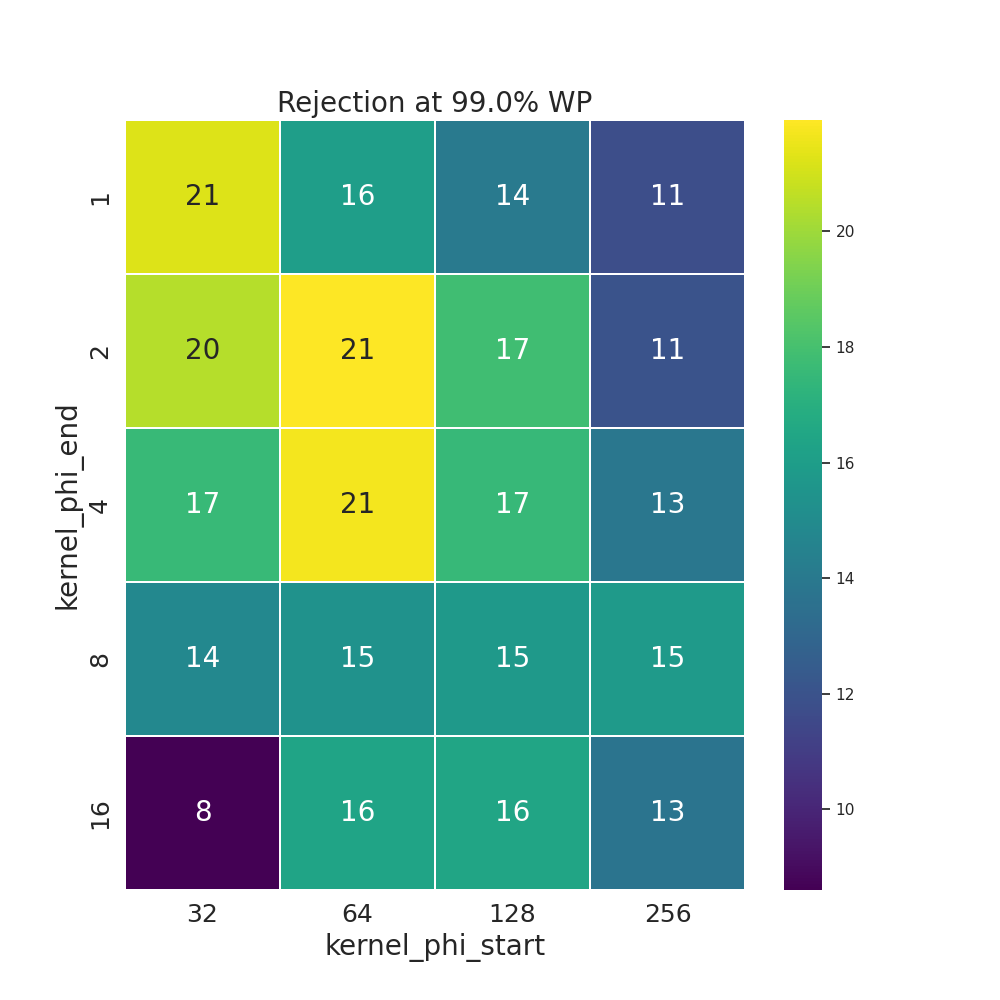}
\end{minipage}
\begin{minipage}[b]{0.45\textwidth}
\centering
\includegraphics[width=\textwidth]{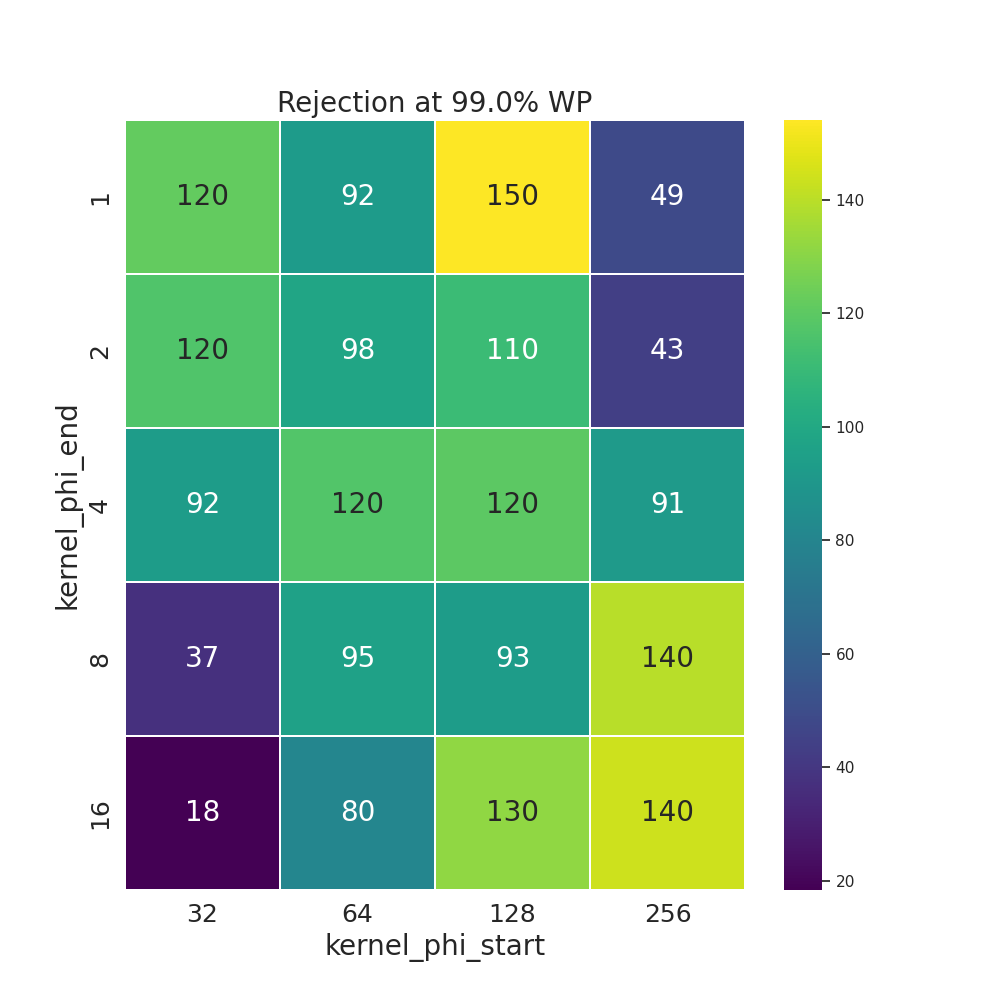}
\end{minipage}
\\
\begin{minipage}[b]{0.45\textwidth}
\centering
\includegraphics[width=\textwidth]{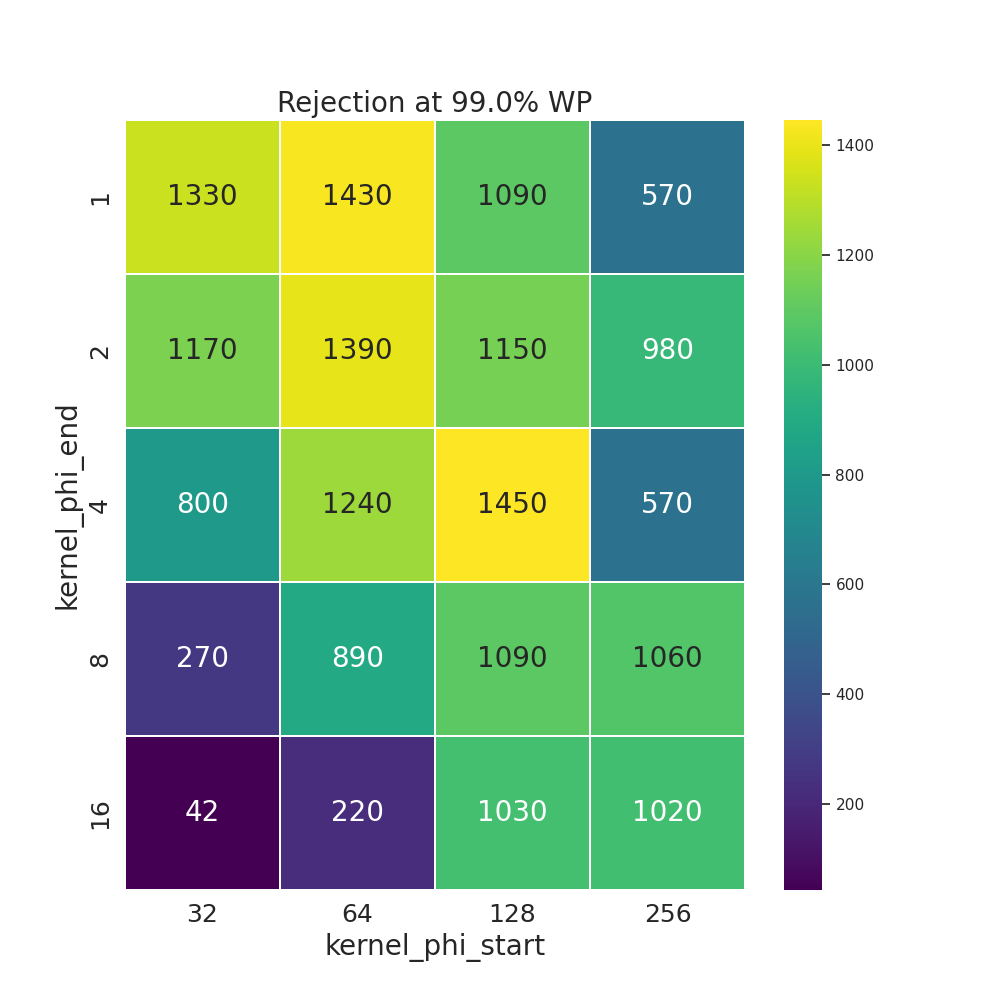}
\end{minipage}
\caption{Background rejection factors when selecting a cut on the output score corresponding to an integrated signal efficiency of~99\% for the hyperparameter scan of the $K_{\text{start}}$ and $K_{\text{end}}$ as described in the text. The three matrices corresponds to the three datasets with $\Delta\varphi=0.001~\text{rad}$ in the top-left, $\Delta\varphi=0.0005~\text{rad}$ in the top-right and $\Delta\varphi=0.0002~\text{rad}$ in the bottom part of the figure.}
\label{fig:kernel_phi_heatmaps}
\end{figure}

The chosen figure of merit for comparing the different trained models is the background rejection for a selection cut scoring an integrated signal efficiency of~99\%. This choice reflects the desire of losing minimal information when filtering in order to not compromise the accuracy when reconstructing tracks. The rejections for the three input datasets, corresponding to $\Delta\varphi=0.001, 0.0005$ or $0.0002~\text{rad}$, and for the varying $K_{\text{start}}$ and $K_{\text{end}}$ is depicted in Figure~\ref{fig:kernel_phi_heatmaps}.

The hyperparameter scans in Figure~\ref{fig:kernel_phi_heatmaps} reveal significant variations in background rejections at the chosen working point and clearly indicate that a finer binning along the $\varphi$ direction leads to improved performance. Among the tested configurations, the model trained with $K_{\text{start}} = 128$ and $K_{\text{end}} = 4$ on the dataset with $\Delta\varphi=0.0002~\text{rad}$ provides the best results and is therefore selected for the subsequent performance and robustness studies, and will henceforth be referred to as the best model.
While lower values for the $\varphi$ binning could be studied, the adopted binning reflects the trade-off among, on one side, the algorithm performance and, the other side, practical constraints related to the dataset size and the number of trainable parameters.
A summary of the dataset, input features, architecture and training configuration is given in Table~\ref{tab:CNN_details}.

\begin{table}[h]
\centering
\begin{tabular}{ll}
\hline
\multicolumn{2}{l}{\textit{Dataset}} \\
Total events & 1\,000\,000 \\
Training / Validation / Test split & 70\% / 15\% / 15\% \\
Signal & Single track, $p_{\text{T}} \in [20, 50]$\,GeV \\
Background & Pile-up tracks and detector noise \\
\hline
\multicolumn{2}{l}{\textit{Input features}} \\
Input representation & 2D hitmap with shape ($\phi$ bins, detector layers) \\
$\Delta\varphi$ binning & 0.0002 \\
$\varphi$ bins & 1000 \\
Detector layers & 8 \\
\hline
\multicolumn{2}{l}{\textit{Architecture}} \\
Model type & CNN-based autoencoder \\
$K_{\text{start}}$ & 128 \\
$K_{\text{end}}$ & 4 \\
$K_{\text{layer}}$ & 6 \\
Trainable parameters & $\sim$120k \\
\hline
\multicolumn{2}{l}{\textit{Training}} \\
Optimizer & Adam \\
Loss function & Binary Cross-Entropy \\
Maximum epochs & 500 \\
Early stopping patience & 20 epochs \\
\hline
\end{tabular}
\caption{Summary of the CNN model training configuration.}
\label{tab:CNN_details}
\end{table}

\begin{figure}[t!]
\centering
\includegraphics[scale=0.8]{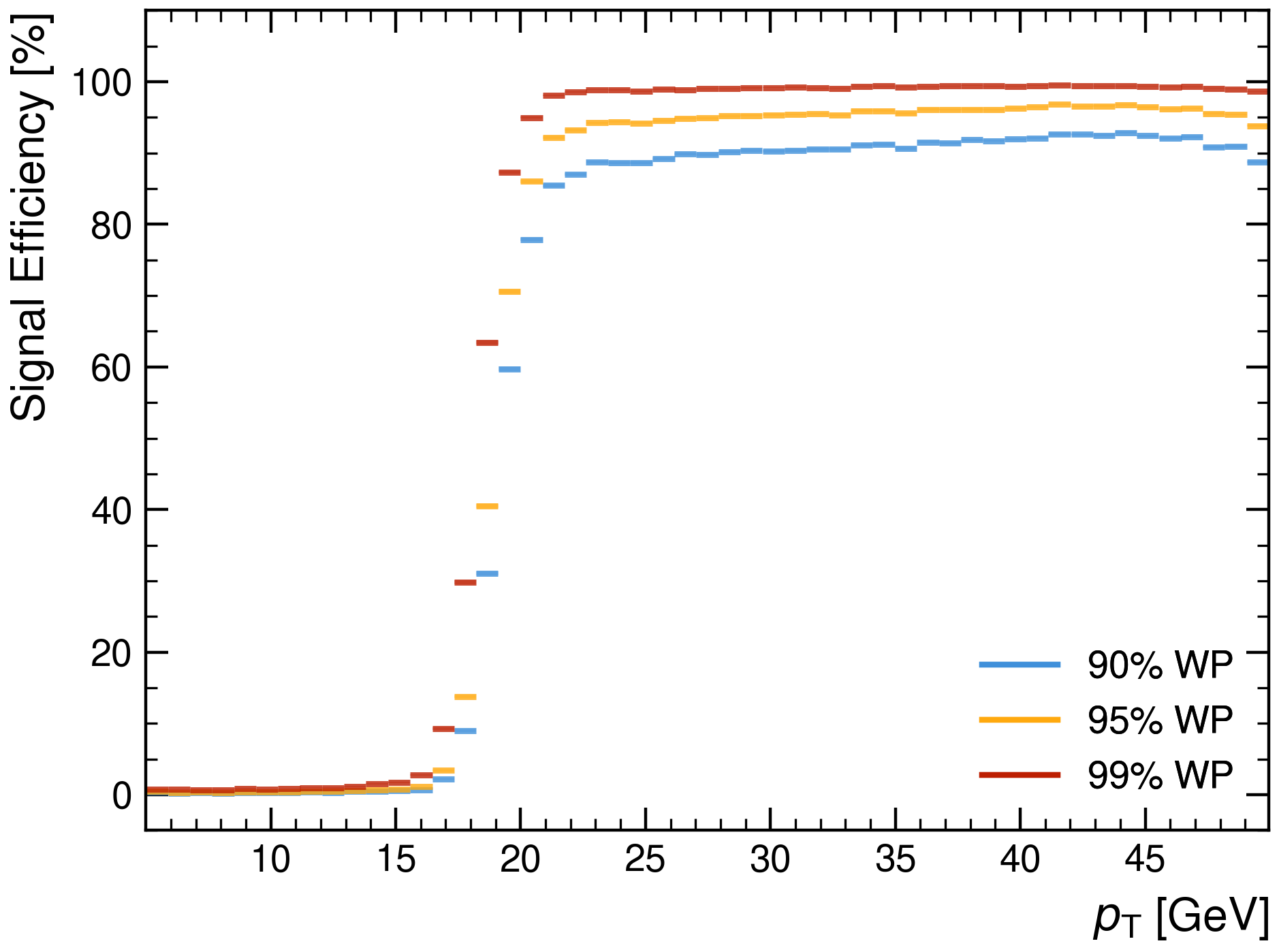}
\caption{Efficiency as a function of $p_{\text{T}}$ of the signal track in the event, computed for the 90\%, 95\% and 98\% WPs, defined as in the text. The error bars are present but not visible, given the large statistics of the test sample.}
\label{fig:pT}
\end{figure}

By defining the filtering efficiency as the ratio between the signal hits correctly classified by the network and the total number of signal hits in the event, its dependence on the signal-track $p_{\text{T}}$ is evaluated on an additional dataset where the signal track is simulated in a $p_{\text{T}}$ range starting at 5~GeV. Working points (WPs) are established using signal tracks in the $p_{\text{T}}$ range between 20 and 50 \text{~GeV}, and a threshold cut on the output score is determined for obtaining average efficiencies of 90\%, 95\%, and 98\%. The efficiency as a function of $p_{\text{T}}$ for the established WPs is shown in Figure~\ref{fig:pT}. As expected, there is no efficiency for low-$p_{\text{T}}$ tracks while the plateaus are at the expected integrated efficiency values for the various WPs. The turn-on is around 20~GeV, as expected considering the kinematics in the training dataset, and only few GeV wide. 

\section{Robustness Tests}\label{Sec:Robustness}
The robustness of the trained network is assessed by evaluating its classification performance on dedicated samples in which increased pile‑up, enhanced hit smearing, and reduced signal‑hit collection efficiency are simulated. The following sub‑sections examine these three aspects individually. This procedure is intended to quantify how the performance scales with rising event complexity and to provide an indication of the importance of incorporating realistic detector effects into the training data.

\subsection{Pile-up}
While the training process was performed on events generated by overlaying an average of 25 pile‑up interactions on top of the additional random noise, the performance of the network is now evaluated on samples in which the average number of secondary collisions is set to 50 and 100. In all cases, the actual distribution of secondary collisions is simulated as Gaussian with a width of $\pm 5$.

\begin{figure}[t!]
\centering
\begin{minipage}[b]{0.45\textwidth}
\hspace{1.5cm}
\includegraphics[width=0.8\textwidth]{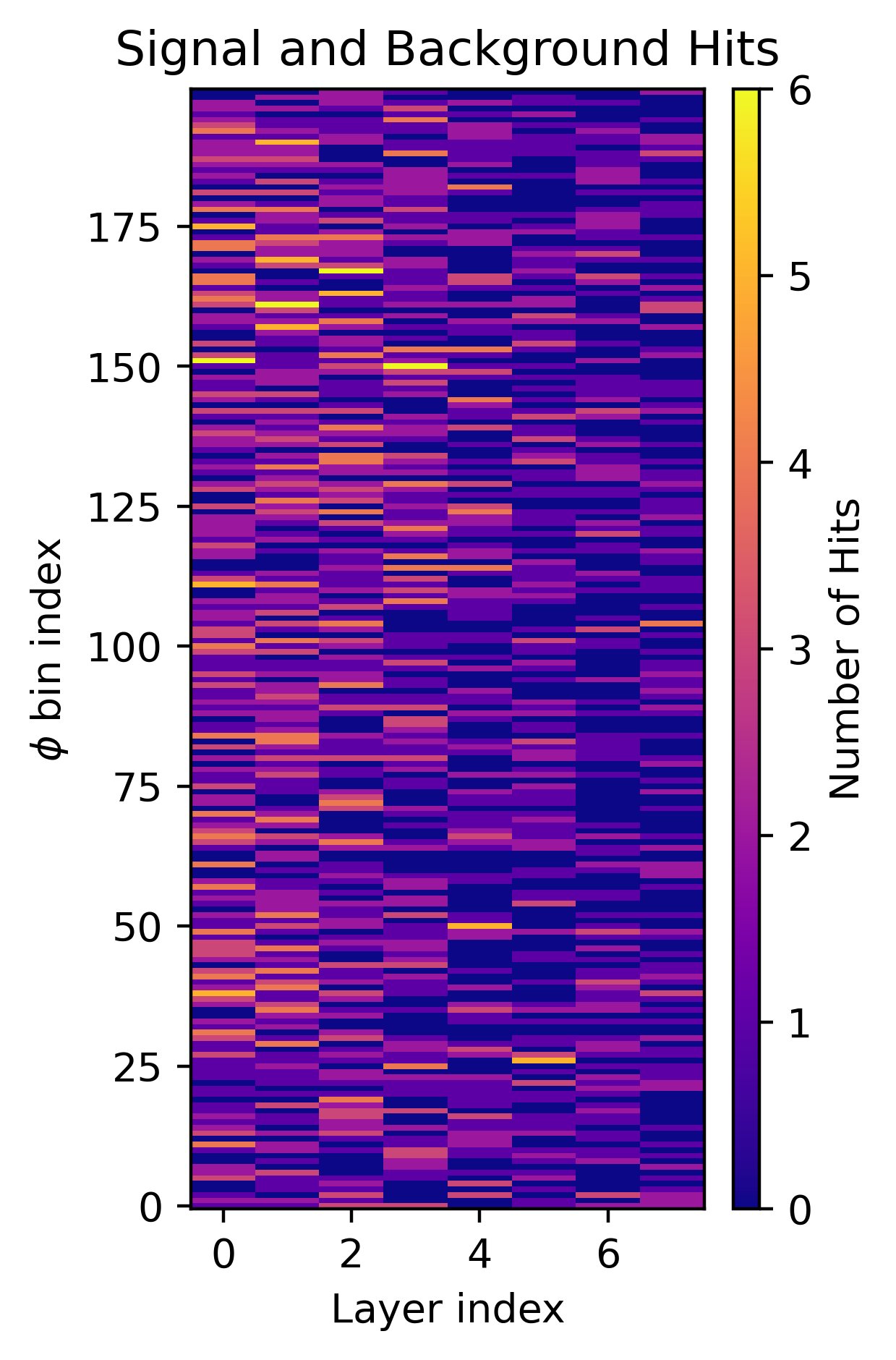}
\end{minipage}
\begin{minipage}[b]{0.45\textwidth}
\includegraphics[width=0.8\textwidth]{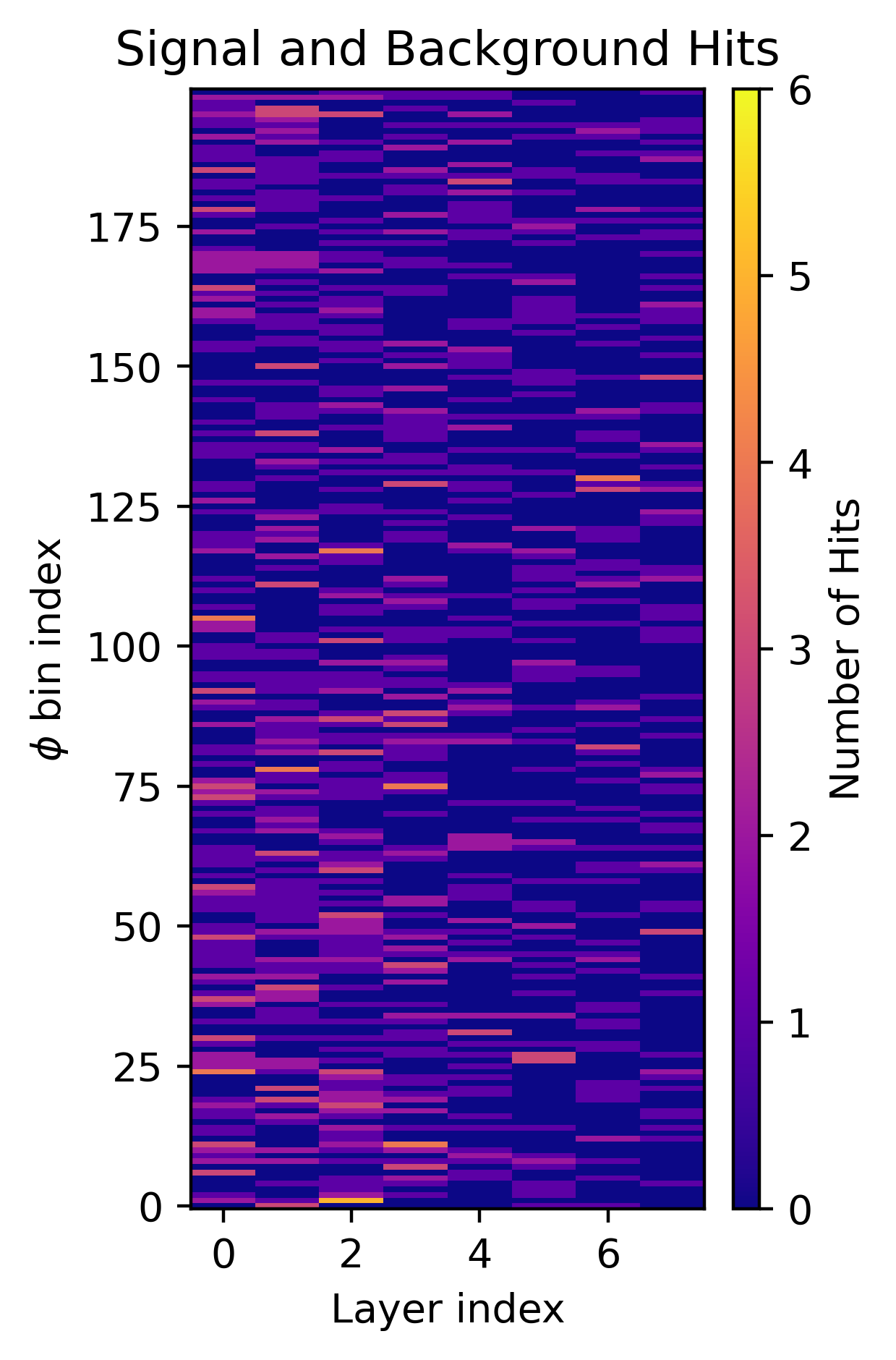}
\end{minipage}
\caption{Displays of the hitmaps as 2D projections in the $(\varphi \text{ bin index}, \text{layer index})$ coordinates as in Figure~\ref{fig:event_image} for an event with 50 pile-up interactions on the left and 100 on the right. Hitmaps are displayed for signal and background, where the signal is a single track from the hard-scattering collision. The binning on the $\varphi$ coordinate is $0.001~\text{rad}$.}
\label{fig:pu_displays}
\end{figure}

Figure~\ref{fig:pu_displays} shows the displays of two events, the left one with an average of 50 pile‑up interactions and the right one with 100. Both the hit density and the number of overlapping hits per bin visibly increase when going towards higher pile-up scenarios.

\begin{figure}[t!]
\centering
\includegraphics[scale=0.8]{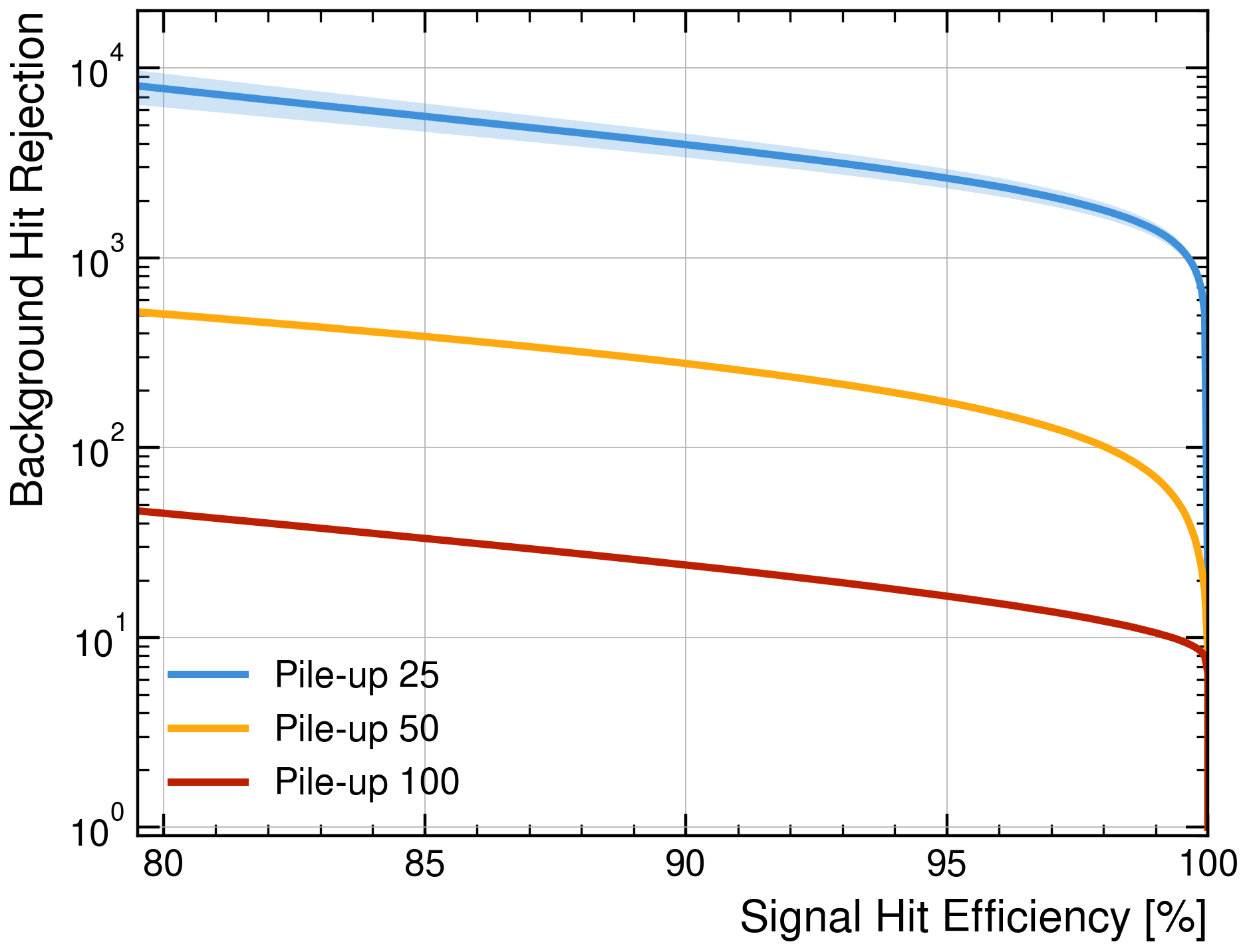}
\caption{ROC curves when the best model is evaluated against events with a higher number of pile-up collisions.}
\label{fig:robustness_pileup}
\end{figure}

The best model is evaluated on the these two simulated samples. The per-hit background rejection versus signal efficiency is shown in Figure~\ref{fig:robustness_pileup}. 

As expected the network performance degrades for higher pileup samples, with the rejection being reduced by a factor $\sim10$ and $\sim100$ for the samples with an average pile-up of 50 and 100, respectively. Even though the network was not exposed to such high hit densities during training, the features learned generalize well enough to allow reliable identification of high‑$p_{\text{T}}$ tracks in considerably more challenging conditions.

\subsection{Smearing}
In generating the datasets described in Section~\ref{Sec:Data}, a smearing of the hit positions, as reported in Table~\ref{tab:smearing}, is applied according to the spatial dimensions of the sensors of the pixel and strip detectors of the ATLAS experiment.

\begin{figure}[htbp]
\centering
\includegraphics[scale=0.8]{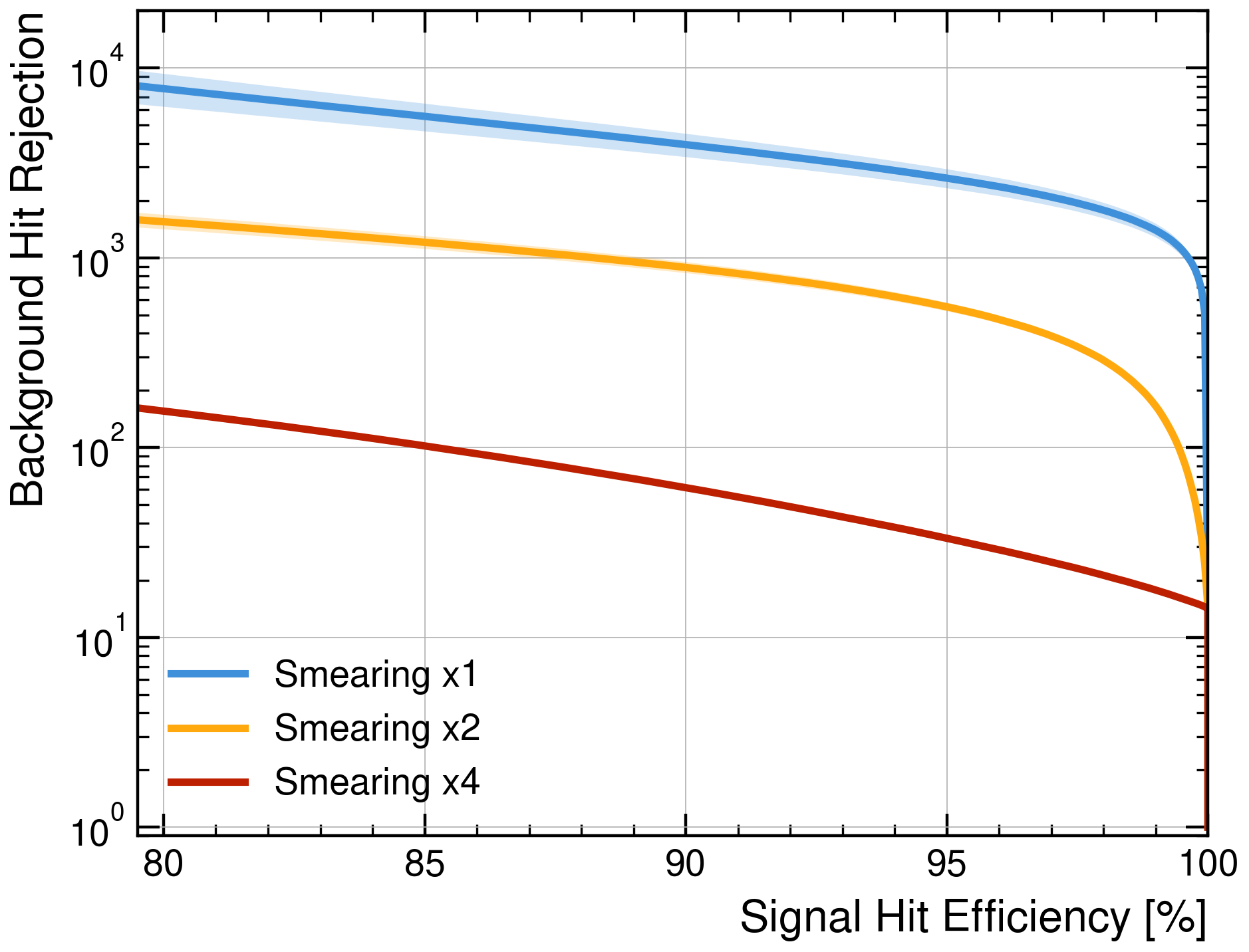}
\caption{ROC curves for the best model evaluated on samples where the Gaussian smearing in the $\varphi$ coordinate of the generated hits is inflated by factors of 2 and 4.}
\label{fig:robustness_smearing}
\end{figure}

The performance of the best model is evaluated with the smearing on $\varphi$ increased by factors of 2 and 4. The $z$ coordinate, instead, is left with the standard smearing, being the input hitmaps a projection of the $x$-$y$ plane. The per-hit background rejection versus signal efficiency is show in Figure~\ref{fig:robustness_smearing}.

The network performance is found to degrade for samples with increased smearing, with the rejection reduced by approximately 8 and 80 for the cases where the smearing in $\varphi$ is increased by factors of 2 and 4, respectively. Although a reduction in rejection power is expected under these harsher conditions, the 99\% working point on the $\times$4‑smearing sample still achieves a rejection larger than 10, demonstrating the robustness of the network against such increased smearing levels.
 
\subsection{Signal Efficiency}
\begin{figure}[htbp]
\centering
\includegraphics[scale=0.8]{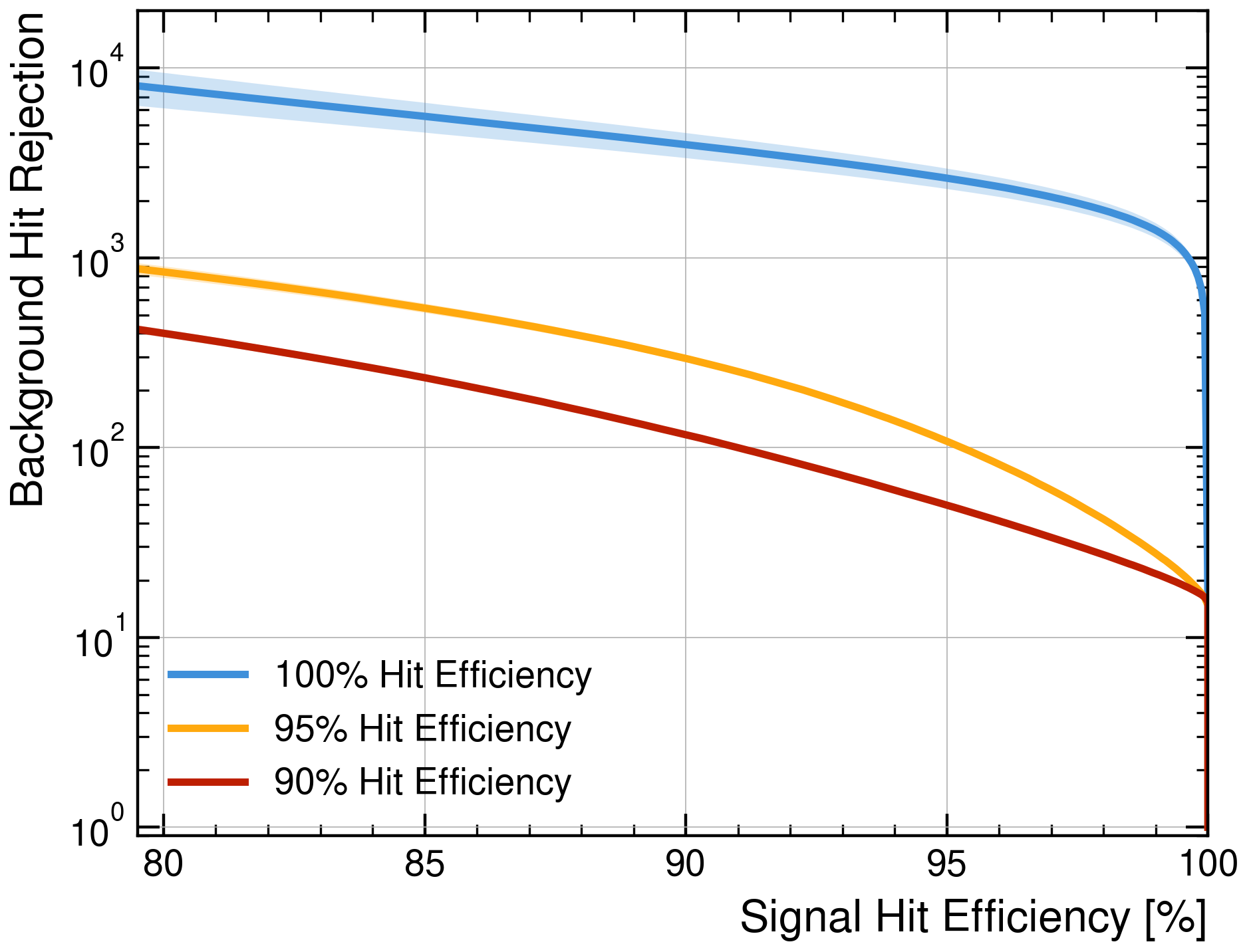}
\caption{ROC curves for the best model evaluated on samples where the per-hit signal efficiency is reduced to 95\% and 90\%.}
\label{fig:robustness_efficiency}
\end{figure}

The performance of the network is also evaluated under reduced signal‑hit efficiency, introducing the possibility of missing hits in the signal tracks due to detector inefficiency. To this end, the best model is tested on dedicated samples in which the signal efficiency in the event generation is set to 95\% and 90\%. Figure~\ref{fig:robustness_efficiency} shows the comparison between the rejections obtained for the nominal efficiency of 1 and for efficiencies of 95\% and 90\%.

A clear reduction in performance is observed, with the ROC curves for the 90\% and 95\% signal‑efficiency samples converging toward similar rejection values as the signal efficiency approaches 1. In particular it is noted that a significant level of background rejection is still maintained even at very high signal efficiencies and is visible in both Figures~\ref{fig:robustness_smearing} and~\ref{fig:robustness_efficiency}. This behaviour is attributed to isolated background hits, which are easily classified by the model and give rise to the peak at zero in the raw $D_{\text{out}}$ distribution shown in Figure~\ref{fig:dicr_output}.

\section{Conclusions and outlook}
This work presents a novel machine learning approach based on a convolutional neural network to classify hits in a simulated tracking detector, distinguishing those produced by charged‑particles from the hard-scattering primary vertex from those produced by pile-up tracks originating from secondary collisions and detector noise. The method addresses the computational challenges of track reconstruction at the trigger level by significantly reducing the number of reconstructed hits to be considered for pattern recognition, and its remarkable performance indicates that it could provide a viable solution for the demanding conditions expected at the future High‑Luminosity LHC. Owing to its simple architecture and relatively small number of parameters, such a filtering algorithm shows promising potential for fast, low‑resource applications.

Future developments include further refinement of the network architecture, the exploration of three-dimensional hitmap representations and their impact on both performance and resource usage, the algorithm development and testing with full-simulation data, and the deployment of the model on GPU and FPGA accelerator cards. A further step will be the integration of the filtering algorithm into a full trigger‑level tracking chain to assess its impact on the overall system performance. 

Several strategies can be considered to integrate this algorithm in a tracking-chain, starting with the straight-forward approach where only the subset of hits corresponding to signal bins is provided to the pattern recognition stage; in this way the pattern recognition step would speed up dramatically, at the cost of throwing away the information related to background bins. An alternative idea is to attach the output from the hit-filtering CNN to each hit, and use this score as a weight for the pattern-recognition step.

Finally, also an hybrid approach can be considered, where the hits with a score below a decided threshold are directly thrown away and the remaining ones are passed to the tracking algorithms together with the score of the hit-filtering CNN.

\section{Acknowledgments}
This work was partially supported by ICSC -- Centro Nazionale di Ricerca in High Performance Computing, Big Data and Quantum Computing, funded by European Union -- NextGenerationEU.


\newpage
\begin{thebibliography}{99}
\bibitem{ref:LHC}
L. Evans and P. Bryant, LHC Machine, JINST {\bf 3} (2008) S08001.
\bibitem{ref:ATLAS-Paper}
ATLAS Collaboration, The ATLAS Experiment at the CERN Large Hadron Collider, JINST {\bf 3} 2008, S08003.
\bibitem{ref:CMS-Paper}
CMS Collaboration, The CMS experiment at the CERN LHC,JINST {\bf 3}(2008), S08004.
\bibitem{ref:ATLAS-Trigger-Run2}
ATLAS Collaboration, Operation of the ATLAS trigger system in Run 2, JINST {\bf15} (2020) 10, P10004.
\bibitem{ref:ATLAS-Trigger-Run3}
ATLAS Collaboration. The ATLAS Trigger System for LHC Run 3 and Trigger performance in 2022, JINST {\bf 19} 2024, P06029.
\bibitem{ref:CMS-Trigger}
CMS Collaboration, The CMS trigger system, JINST {\bf 12} 2017, P01020.
\bibitem{ATLAS:2021lws}
ATLAS Collaboration, The ATLAS inner detector trigger performance in pp collisions at 13~TeV during LHC Run 2, Eur. Phys. J. C \textbf{82} (2022) no.3, 206.
\bibitem{ref:HL-LHC}
O. Aberle, \textit{et al.}, High-Luminosity Large Hadron Collider (HL-LHC): Technical design report, CERN-2020-010 (2020).
\bibitem{Duarte:2020ngm}
J.~Duarte and J.-R.~Vlimant,
Graph Neural Networks for Particle Tracking and Reconstruction,
Artificial Intelligence for High Energy Physics,
pp.~387--436 (2022), \url{https://doi.org/10.1142/9789811234033_0012}
\bibitem{DeZoort:2021rbj}
G.~DeZoort, \textit{et al.}, Charged Particle Tracking via Edge-Classifying Interaction Networks,
Comput. Softw. Big Sci. \textbf{5} (2021), 26.
\bibitem{Bocci:2020pmi}
A.~Bocci, , \textit{et al.}, Heterogeneous Reconstruction of Tracks and Primary Vertices With the CMS Pixel Tracker, Front. Big Data \textbf{3} (2020), 601728.
\bibitem{ATLAS-HL-LHC-Computing}
ATLAS Collaboration, ATLAS Software and Computing HL-LHC Roadmap, CERN-LHCC-2022-005.
\bibitem{CMS-HL-LHC-Computing}
CMS Collaboration, CMS Phase-2 Computing Model: Update Document, CERN-CMS-NOTE-2022-008.
\bibitem{Coccaro:2023nol}
A.~Coccaro, \textit{et al.}, Fast neural network inference on FPGAs for triggering on long-lived particles at colliders, Mach. Learn. Sci. Tech. \textbf{4} (2023) no.4, 045040.
\bibitem{Soybelman:2024mbv}
N.~Soybelman, \textit{et al.}, Accelerating graph-based tracking tasks with symbolic regression, Mach. Learn. Sci. Tech. \textbf{5} (2024) no.4, 045042.
\bibitem{ATLAS:IBL_TDR}
ATLAS Collaboration, M.~Capeans \textit{et al.},
ATLAS Insertable B-Layer Technical Design Report,
CERN-LHCC-2010-013, ATLAS-TDR-19 (2010), \url{https://cds.cern.ch/record/1291633}
\bibitem{ATLAS:Pixel_TDR}
ATLAS Collaboration, G.~Aad \textit{et al.},
ATLAS pixel detector electronics and sensors,
JINST {\bf 3} (2008) P07007, \url{https://doi.org/10.1088/1748-0221/3/07/P07007}.
\bibitem{ATLAS:SCT_precision}
ATLAS Collaboration,
Alignment Performance of the ATLAS Inner Detector Tracking System in 7 TeV proton-proton collisions at the LHC,
ATLAS-CONF-2010-067 (2010).
\bibitem{ref:ATLAS-PhaseII}
ATLAS Collaboration, Technical Design Report for the ATLAS Inner Tracker Pixel Detector, Technical design report, CERN-2017-21 (2017).
\bibitem{ATLAS:2024rnw}
ATLAS Collaboration, Expected tracking performance of the ATLAS Inner Tracker at the High-Luminosity LHC,
JINST \textbf{20} (2025) 02, P02018.
\bibitem{ATLAS:2016zkp}
ATLAS Collaboration, Charged-particle distributions in $\sqrt{s}$ = 13 TeV pp interactions measured with the ATLAS detector at the LHC, Phys. Lett. B \textbf{758} (2016), 67-88.
\bibitem{ref:reduce_dim} 
G.~E.~Hinton, \textit{et al.}, Reducing the Dimensionality of Data with Neural Networks, Science, 313 (5786): 504-507 (2006).
\bibitem{ref:denoise_ae} 
V.~Pascal, \textit{et al.}, Extracting and composing robust features with denoising autoencoders,
ICML, ACM International Conference Proceeding Series, 1096-1103 (2008).
\bibitem{ref:segmentation} 
J.~Long, \textit{et al.}, Fully convolutional networks for semantic segmentation, 2015 IEEE Conference on Computer Vision and Pattern Recognition, 3431-3440 (2015).
%\bibitem{ref:alignment}
%ATLAS Collaboration, Alignment of the ATLAS Inner Detector in Run-2, Eur. Phys. J. C \textbf{80} (2020) no.12, 1194.
\end{thebibliography}
\end{document}